\documentclass[twocolumn,english,aps,prb,floatfix,superscriptaddress]{revtex4}
\usepackage[T1]{fontenc}
\usepackage[latin9]{inputenc}
\usepackage{bm}
\usepackage[pdftex]{graphicx}
\usepackage{esint}
\usepackage{subfigure}
\usepackage{tikz}
\usepackage{pgfplots}

\makeatletter

\newcommand{\lyxdot}{.}

\@ifundefined{textcolor}{}
{%
 \definecolor{BLACK}{gray}{0}
 \definecolor{WHITE}{gray}{1}
 \definecolor{RED}{rgb}{1,0,0}
 \definecolor{GREEN}{rgb}{0,1,0}
 \definecolor{BLUE}{rgb}{0,0,1}
 \definecolor{CYAN}{cmyk}{1,0,0,0}
 \definecolor{MAGENTA}{cmyk}{0,1,0,0}
 \definecolor{YELLOW}{cmyk}{0,0,1,0}
}

\usepackage{psfrag}

\makeatother

\usepackage{babel}

\begin{document}

\title{Electrostatic effects and band-bending in doped topological insulators}

\author{Dimitrios Galanakis}
\affiliation{School of Physical and Mathematical Sciences, Nanyang Technological University, Singapore 637371}
\author{Tudor D. Stanescu}
\affiliation{Department of Physics, West Virginia University, Morgantown, WV 26506}

\begin{abstract}
We investigate the electrostatic effects in doped topological insulators by developing a self consistent scheme for an interacting tight binding model.
The presence of bulk carriers, in addition
to surface electrons, generates an intrinsic inhomogeneous charge density
in the vicinity of the surface and, as a result, band bending effects are present. 
We find that electron doping and hole doping produce band bending effects of 
similar magnitude and opposite signs. The presence of additional surface
dopants breaks this approximate electron-hole symmetry and dramatically 
affects the magnitude of the band bending. Applying a gate potential can generate a depletion zone characterized by a vanishing carrier density. We find that the density profile in the transition zone between the depleted region and the bulk is independent of the applied potential.
In thin films the electrostatic effects are 
strongly dependent on the carrier charge density. In addition, we find 
that substrate induced potentials can generate a Rashba type spin-orbit 
coupling in ultra thin topological insulator films. We calculate the profiles of bulk and 
surface states in topological insulator films and identify the conditions corresponding to both types of states being
localized within the same region in space.

\end{abstract}
\maketitle

\section{Introduction}

The existence of metallic surface states with Dirac--like dispersion represents the hallmark of topological insulators.\cite{Kane2005,Bernevig2006,Fu2007a, Fu2007b,Moore2007,Roy2009}  In three dimensional (3D) topological insulators (TIs), the partially occupied surface states form a helical metal that has been predicted to host a wide range of new physical phenomena.\cite{Qi2008,Essin2009,Fu2008,Seradjeh2009} Signatures of characteristic TI surface states have been observed in a family of strongly spin--orbit interacting  $Bi$--based materials.\cite{Hsieh2008,Xia2009,Hsieh2009} However, most of the systems studied experimentally are not three dimensional topologically ordered\cite{Wen1990} bulk solids, but rather doped TIs. Realizing bulk insulating samples, as well as  separating the surface and bulk contributions in various types of measurements, represent serious challenges. 

In doped 3D topological insulators, surface carriers that occupy
the gapless surface states coexist with bulk carriers. As a result,
the charge density in the vicinity of the TI surface is intrinsically
non-homogeneous. Consequently, electrostatic effects, including band
bending in the vicinity of the surface, contribute significantly to
the low-energy physics in these systems. Additional electrostatic
effects may be generated by external electric fields that are present
in various experimental setups involving TIs. Examples include applied
gate potentials that create a bulk depletion zone and expose the surface
states, gate potentials that control the bulk carrier density in thin
films, and substrate-induced potentials that modify the band structure
in epitaxially-grown films.

In this paper we present a systematic theoretical study of electrostatic
effects in doped TIs using an effective four band tight binding model\cite{Hutasoit2011}
for the Bi-family topological insulators. The Coulomb interaction
is included at the Hartree level using a self-consistent scheme. We
focus on systems with a slab geometry with $(111)$ surfaces. This
surface orientation is the most relevant for experiment and is characterized
by topological surface states with a localization length scale of
the order of $1$nm. Note that other surface orientations are characterized
by surface states with much larger localization length scales and,
consequently, will exhibit significantly different electrostatic effects.
The slab geometry allows us to study the effects of uniformly distributed
bulk and surface dopants, e.g., those originating from the presence
of $Se$ vacancies. Our calculations ensure a realistic treatment of the density profiles originating from
topological surface states, which are essential for understanding
bulk screening effects and cannot be described within a simple Thomas-Fermi
approximation. The specific problems that we address have directly
measurable consequences and we explicitly emphasize the links between
our findings and experiment.

\vspace{-5mm}

\subsubsection*{Structure of the paper and main results}

In section II we describe the tight--binding model that we use for characterizing the low-energy properties of the doped TIs and the self--consistent method that we implement numerically to account for the effects of Coulomb interaction.  Our main results are presented in section III (for bulk TIs) and section IV (for TI thin films). Section V contains our conclusions.

In Section III we investigate the electrostatic effects that occur in the vicinity of the surface of a bulk TI or near the interface between a TI and a trivial insulator. First, we address the question concerning the accuracy of Thomas-Fermi approximation in low-doped TIs (section III A). We find that the Thomas-Fermi approximation is consistent with the numerical self-consistent calculations in regions with bulk charge density, but fails to describe the depleted region that may occur near the surface, as well as surface charge density contributions. Next, we study band bending in the vicinity of TI surfaces (section III B). The presence of this effect in TIs has direct experimental consequences, e.g., it generates  differences\cite{Analytis2010} between surface-sensitive probes, such as the angle-resolved photoemission spectroscopy, and bulk-sensitive probes, for example  Shubnikov--de Haas measurements. In doped TIs, band bending is the result of the bulk electrons being pushed away by the surface charge of the helical metal. In addition, surface dopants, such as Se vacancies\cite{Xia2009,Hsieh2009} that migrate to the surface, can significantly modify the bending of the bands at the boundary. We investigate the difference between band bending in electron--doped and hole--doped TIs, as well as  the role of surface dopants. To facilitate visualizing the band bending effects, we calculate the local density of states (LDOS) in the vicinity of the surface. Band bending, together with the charge density contributions of surface states, generate intrinsically  non-homogeneous density profiles and  effective electrostatic potentials that vary significantly in the vicinity of the system boundaries. These potentials can be probed experimentally, for example using second harmonic generation.\cite{Hsieh2011,Hsieh2011a}
We also study the charge  density profiles in the presence on an external field. We find that applying a gate potential can create a depletion zone near the surface characterized by a vanishing bulk carrier density, while deep inside the bulk the potential is screened and the charge density is constant. We also find that the density profile in the transition zone between the depleted region and the bulk is independent of the applied potential. Creating a depleted region could allow access to the surface states without interference from the bulk carriers.  

In section IV we investigate different types of electrostatic effects that can occur in topological insulator thin films. In general, screening in TI films is significantly weaker that in bulk systems. Consequently, the band bending effects can be quite severe and, typically, the system does not  have a bulk region characterized by constant charge density. First, we study the effects of an external potential, which may occur, for example, as a result of a charge transfer from the substrate\cite{Zhang2010} (section IV A). We find that the substrate-induced band bending is strongly dependent on the carrier density. In addition, in ultra--thin TI films a gap  opens at the Dirac point as a result of the overlap between the surface states from the two surfaces and,  in the presence of a substrate--induced potential difference, an effective Rashba-type spin orbit splitting is generated.\cite{Zhang2010} Finally, we calculate the profiles of the surface and bulk states in TI films  (section IV B)  and show that, in certain conditions, the lowest energy states from the conduction band become localized near the TI surface. This opens the interesting possibility of coupling surface and bulk states, which typically are spatially separated, using, for example, optically--induced transitions.

\section{Model and method}

Topological insulators from the Bi family have a layered structure consisting in quintuple layers of Bi and Se (or Te) atoms oriented parallel to the $(1 1 1)$ plane. In a slab with $(1 1 1)$ surfaces,  the potential created by electrons
occupying the topological surface states is screened by the bulk charges.
To study the resulting charge density profile we consider a minimal
model described by the Hamiltonian:
\begin{equation}
H_{MF}=H_{TI}+H_{int}.\label{eq:MF_Hamiltonian}
\end{equation}

The first term, $H_{TI}$, is an effective tight-binding model on
a triangular lattice with 4-bands per quintuple layer,

\begin{equation}
H_{TI}=\sum_{\alpha ij}\left(\epsilon_{0}^{\alpha}\delta_{ij}+t_{ij}^{\left(\alpha\right)}\right)\bm{c}_{i\alpha}^{\dagger}\bm{c}_{j\alpha}+\bm{c}_{i\alpha}^{\dagger}\left(i\lambda_{ij}\bm{\delta}\cdot\hat{\bm{\sigma}}\right)\bm{c}_{j\overline{\alpha}}.\label{TI_Hamiltonian}
\end{equation}
where $\bm{c}_{i\alpha}^{\dagger}=\left(f_{i\alpha\uparrow}^{\dagger},f_{i\alpha\downarrow}^{\dagger}\right)$
are spinors and $f_{i\alpha\sigma}^{\dagger}$ creates an electron
in one of the 4 bands labeled by $\alpha=1,2$ and $\sigma=\uparrow,\downarrow$.
The first and the second term of $H_{TI}$ describe the intra and
inter band hoping respectively. The tight binding model parameters used in this calculation are
the intra layer hoping elements $t_i^{(1)}=1.42995$ and $t_i^{(1)}=0.0318196$, the inter-layer hoping elements
 $t_o^{(2)}=-2.95299$ and $t_o^{(2)}=-0.0413289$ and band edges
$\epsilon_0^{(1)}=-8.527$, $\epsilon_0^{(2)}=17.6495$ and the spin orbit couplings 
$\lambda_{(1)}=0.291864$ and $\lambda_{(2)}=0.115223$ all measured in $eV$. 
The Coulomb repulsion, $H_{int}$,
is described at the mean field level by the term, 
\begin{equation}
H_{int}=e^{2}\sum_{i\neq j}\frac{\left(\hat{n}_{i}-n_{ion}\right)\left(\left\langle \hat{n}_{j}\right\rangle -n_{ion}\right)}{\left|{\bf R}_{i}-{\bf R}_{j}\right|},\label{eq:interaction_hamiltonian}
\end{equation}
where $\hat{n}_{i}=\sum_{\alpha,\sigma}f_{i\alpha\sigma}^{\dagger}f_{i\alpha\sigma}$
is the local electron density and $n_{ion}$ is the ionic density. 

In a slab geometry, the in-plane wavevector ${\bf k}_{\parallel}$
is a good quantum number, due to the translational invariance along
the layers. The local density, $\left\langle \hat{n}_{j}\right\rangle $,
depends only on the layer index and can be expressed as the average
of contributions coming from the different momentum sectors. For a
slab of $N$ layers, each sector contains $4N$ orbitals and is diagonalized
independently. For a given density profile, the mean field local chemical
potential can be determined by solving self consistently the discrete
one dimensional Poisson equation,
\begin{equation}
\mu_{l}=\mu_{0,L}+\delta\mu_{l}\left(\left\langle n_{l}\right\rangle \right),\label{eq:Poisson_1D}
\end{equation}
where 
\begin{equation}
\delta\mu_{l}\left(\left\langle n_{l}\right\rangle \right)=2U_{\parallel} \left\langle \delta n_{l}\right\rangle+U_{\perp}\sum_{l^{\prime}=1}^{N}\left|l-l^{\prime}\right|  \left\langle \delta n_{l}\right\rangle
\label{eq:delta_mu_local}
\end{equation}
 vanishes at uniform electric density, $\left\langle \delta n_{l}\right\rangle =\left\langle n_{l}\right\rangle - n_{e}$, $n_{e}$ the total electron density 
and 
\begin{equation}
\mu_{0,l}=\mu_{0}-El-P\left(l-\frac{1+N}{2}\right)^{2}-S_{l},
\end{equation}
is a constant external potential, where the magnitude of the external electric field $E$, 
the parabolic potential $P$ and a surface potential $S_{l}$ are given by
\begin{eqnarray}
E & = & E_{ext}+U_{\perp}\left(\delta n_{surf,1}-\delta n_{surf,N_{L}}\right)\\
P & = & U_{\perp}\left(n_{ion}-n_{e}\right)\\
S_{l} & = & 2U_{\parallel}\left[\delta_{l,1}\delta n_{surf,1}+\delta_{l,N_{L}}\delta n_{surf,N_{L}}\right].
\end{eqnarray}
 Here $E_{ext}$ represents the external electric field, $\delta n_{surf,1}$,
$\delta n_{surf,N}$ the doping of the first and last surface respectively
measured relative to the ion density, and $U_{\perp}$, $U_{\parallel}$
are respectively the intra and inter layer Coulomb repulsion. The
values of the coupling constants are $U_{\perp}=\frac{2ce^{2}}{\sqrt{3}a^{2}\epsilon\epsilon_{0}}$
and $U_{\parallel}=\gamma\frac{a}{c}U_{\perp}$ where $a$ is the
lattice constant of each layer, $c$ the inter layer distance, $\gamma\approx0.1936$
a geometrical factor and $\epsilon\approx90$  the electric
permitivity of the bulk. This corresponds to a value of the interaction
parameter $U_{\parallel}=4eV$which is the one that we will use throughout
this paper.

\subsection*{Carrier density}

In this paper we will present a self-consistent mean-field calculation of
screening effects in topological insulators that describes the density 
profiles generated by the contributions  of both surface and bulk states.
In a slab with $N_{L}$ layers the carrier density is a sum of two contributions,
the bulk doping and the surface doping,
\begin{equation}
\delta n_{e}=\delta n_{ion}+\frac{\delta n_{surf,1}+\delta n_{surf,N}}{N_{L}}.\label{eq:total_carrier_density}
\end{equation}
where $\delta n_{ion}$ and $\delta n_{surf}$ is the bulk and surface
doping respectively. The bulk doping typically comes from $Se$ vacancies. Their
volume density, $\delta\rho_{ion}$, varies between $10^{17}-10^{18}cm^{-3}$, 
which translates in a per unit-cell doping of $\delta n_{ion}=V_{U.C.}\delta\rho_{ion}\sim10^{-5}-10^{-4}$,
where $V_{U.C.}\approx1.42\times10^{-22}cm^{-3}$ is the volume of
the unit cell. The surface doping comes from $Se$ vacancies on the
surface created during the cleaving process or donors that come from
the solution in which the sample is stored. Their surface
density $\delta\sigma_{surf}$ typically varies between $10^{12}cm^{-2}$
and $10^{13}cm^{-2}$. This corresponds to a per-surface layer doping
of $\delta n_{surf}=\frac{\sigma_{surf}V_{U.C.}}{c/3}\sim10^{-3}-10^{-2}$,
which can be $10-1000$ times bigger than the bulk doping at the surface
layer.

\begin{figure}
\includegraphics[width=0.9\columnwidth]{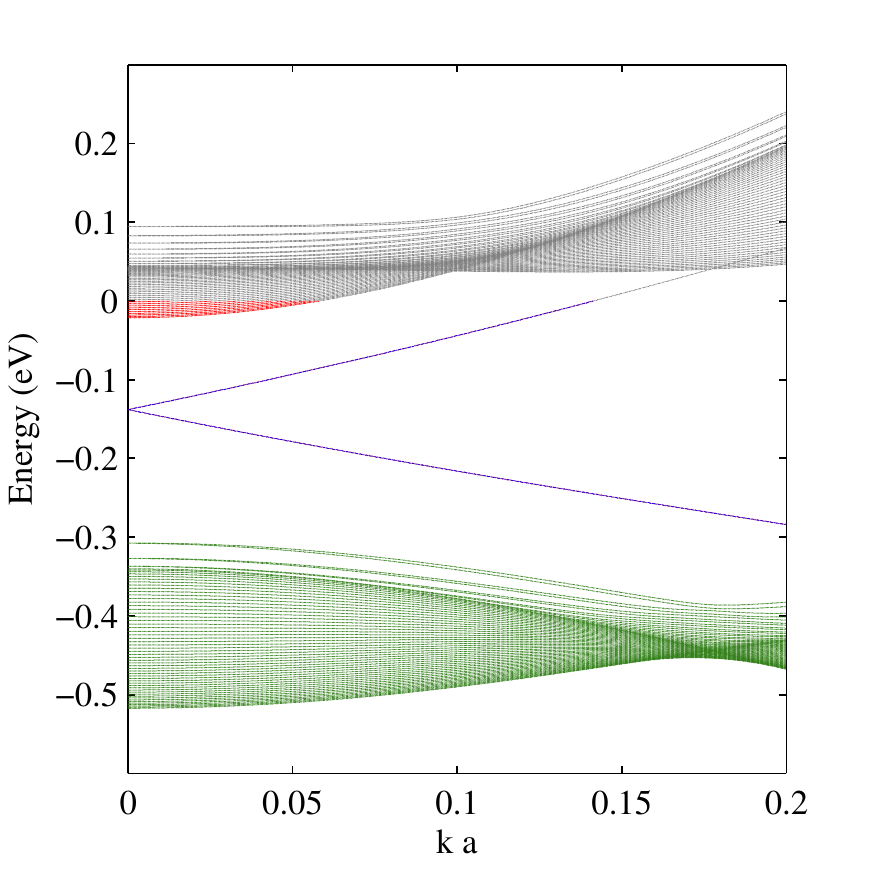}

\caption{(color online) The energy spectrum for $\rho=6\times10^{-4}$  
as a function of $\left|{\bf k}_{\parallel}\right|$,
featuring the occupied (red) and unoccupied (grey) states of the conduction
band, the valence band states (green) and the Dirac cones (purple) which
are separated due to the bias. \label{fig:energy_spectrum}}
\end{figure}

\section{Electrostatic effects near the surface of a bulk topological insulator}

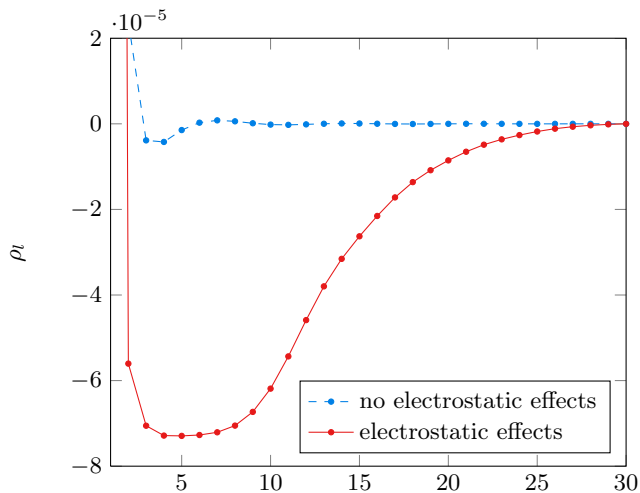
\begin{figure}
%
%
%
\begin{tikzpicture}

\definecolor{mycolor1}{rgb}{0,0.5,0.9}
\definecolor{mycolor2}{rgb}{0.9,0.1,0.1}

\begin{axis}[
xmin=1, xmax=30,
ymin=-8e-05, ymax=2e-05,
ylabel={$\rho_l$},
legend style={at={(0.97,0.03)},anchor=south east,nodes=right}]

\addplot [color=mycolor1,dashed,mark size=1pt,mark=*,mark options={solid}]
coordinates{
 (2,2.38912331740337e-05)(3,-3.86672330954241e-06)(4,-4.23114782854839e-06)(5,-1.45796810802068e-06)(6,2.71630304027326e-07)(7,8.09826814496262e-07)(8,5.85944610609346e-07)(9,1.33125356516217e-07)(10,-1.79333445426977e-07)(11,-2.38742898783784e-07)(12,-1.26308706605016e-07)(13,1.46930472233464e-08)(14,8.68082907778955e-08)(15,7.50841944174852e-08)(16,2.1317752896266e-08)(17,-2.42298998642809e-08)(18,-3.65932293355797e-08)(19,-2.12036717073261e-08)(20,1.43374601080382e-09)(21,1.42888483267711e-08)(22,1.30546995436021e-08)(23,3.97395494289299e-09)(24,-4.11263956223706e-09)(25,-6.39932951074229e-09)(26,-3.73520592233945e-09)(27,2.81783485434062e-11)(28,1.78672809880709e-09)(29,1.18494147827164e-09)(30,0)(31,0) 
};

\addlegendentry{no electrostatic effects};

\addplot [color=mycolor2,solid,mark size=1pt,mark=*,mark options={solid}]
coordinates{
 (1,0.000977289307409368)(2,-5.60347460747401e-05)(3,-7.05455272300703e-05)(4,-7.28493743835656e-05)(5,-7.29196174118485e-05)(6,-7.2715100409404e-05)(7,-7.20813158809896e-05)(8,-7.05196313339762e-05)(9,-6.73124672050385e-05)(10,-6.18766283593253e-05)(11,-5.43474749918893e-05)(12,-4.58632049609164e-05)(13,-3.7979711453584e-05)(14,-3.15502254255584e-05)(15,-2.62845191727479e-05)(16,-2.15301164052661e-05)(17,-1.71961339088256e-05)(18,-1.36196890232476e-05)(19,-1.08455945349029e-05)(20,-8.54425070428988e-06)(21,-6.52456362404408e-06)(22,-4.87595581910583e-06)(23,-3.62919986862664e-06)(24,-2.63586107651648e-06)(25,-1.79396745325633e-06)(26,-1.1343550014864e-06)(27,-6.72523354783294e-07)(28,-3.49991964743168e-07)(29,-1.22921009371169e-07)(30,0)(31,-7.54951656745106e-15) 
};

\addlegendentry{electrostatic effects};

\end{axis}
\end{tikzpicture}

\caption{(color online) The local excess charge density, $\rho_{l}-\rho_{Bulk}$ in the vicinity
of the surface with and without electrostatic effects for a system with  bulk density
$8\times10^{-5}$/cell.  In the absence of electrostatic
effects, the density of bulk carriers in a charge neutral system almost uniform. Small variations are present in a thin surface region consisting in a few quintuple layers. In the presence of electrostatic effects, the charge density associated with surface states pushes the bulk carriers away from the surface creating a depletion zone. As a result,  
the electric field becomes nonzero in wide region that extends more than 20 quintuple layers from the surface.
\label{fig:strength_electrostatic_effects}}
\end{figure}

Topological insulator samples exhibit very rich electrostatic behavior, especially
near the surface. This is mainly due to the presence of bulk impurities, e.g., Se vacancies~\cite{Hsieh2009}, which results in an excess charge in the bulk. The surface itself may contain 
charged impurities, due to air exposure or due to the migration of Se vacancies from the bulk~\cite{Hsieh2009}, 
which provides additional  contributions to the total charge of the system. At the surface, the localized
charge from the topological states repels the bulk charge and, as a result, a partly depleted zone 
zone separates the surface and bulk regions. In the surface region and in the depleted zone the electric field in nonzero.  The presence of electric fields near the surface results
in the bending of the bands upwards or downwards, depending on the sign of the bulk carriers and of the surface dopants. Understanding the detailed band bending mechanism
is essential for the interpretation of various surface and bulk sensitive measurements. In this paper we perform a systematic study of these effects this by solving the electrostatic problem self-consistently.
To emphasize the necessity of a self consistent scheme that accounts for the interaction, we compare 
the density profiles obtained in the presence and in the absence of interactions. 
As shown in Fig. \ref{fig:strength_electrostatic_effects} for a 60 layer slab with bulk 
density of $8\times10^{-5}$/cell, the presence of interactions results in the development 
of a depletion zone near the surface. This non-homogeneous density profile generates a position--dependent electrostatic potential that modifies the spectral properties of the system.

\subsection{Accuracy of Thomas-Fermi approximation}

\begin{figure}
\input{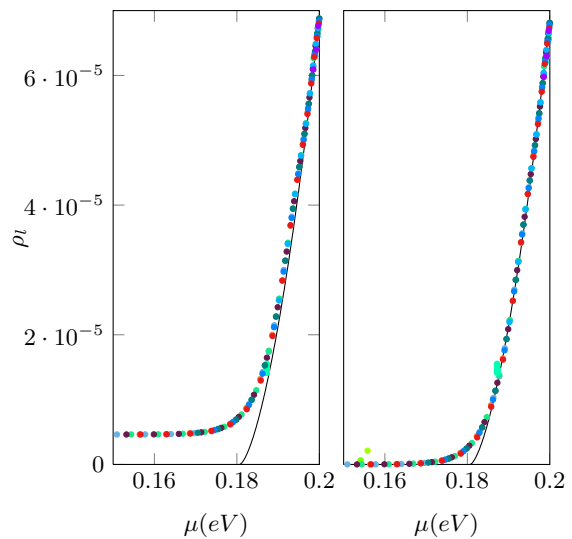}

\caption{(color online) Left:  The self consistent local charge density $\rho_{l}$
plotted against the local $\mu_{l}$, for a 60 layer slab with various densities and external
biases $\Delta V=0-0.4eV$. The solid line is the Thomas-Fermi approximation. The discrepancy
is larger at low densities where there is a residual charge density due to valence band contributions. Right: Conduction band charge density versus local potential. Note that the $\rho_{conduction}$ vs $\mu$ curves are in better agreement
with the Thomas-Fermi approximation everywhere except the tail of
the transition zone.\label{fig:rho_mu_thomas_fermi}}
\end{figure}

In metals and semiconductors screening effects are accurately described in the context
of the Thomas-Fermi (TF) approximation in which the local charge density
$\rho\left({\bf x}\right)$ at position ${\bf x}$ is a function of
the local chemical potential $\mu\left({\bf x}\right)$ 
\begin{equation}
\rho\left({\bf x}\right)\approx\int_{\epsilon_{b}}^{\mu\left({\bf x}\right)}N_{cond}\left(\epsilon\right)d\epsilon,\label{eq:thomas-fermi}
\end{equation}
where $\epsilon_{b}$ is the band edge and $N_{cond}\left(\epsilon\right)$
the density of states of the conduction band. The electrostatic problem
is then reduced in solving the Poisson equation for $\mu\left({\bf x}\right)$.
This approximation is violated at low densities, when the screening
length $\xi$ is much smaller than the de-Broglie wave length of the
conduction electrons. For topological insulators this happens near
the surface, where the topological surface charges repel the bulk
charges to create a charge depletion zone as shown in Fig. \ref{fig:strength_electrostatic_effects}.
Furthermore, the profiles of the surface states extend up to 3-4 layers 
into the bulk and need to be determined explicitly by solving a quantum problem. 
To verify the validity of the Thomas-Fermi approximation, we calculate the density profiles and the local electrochemical potentials for a thick TI slab in the presence of an external bias potential of varying strength and compare the relation between these two quantities with that predicted by the Thomas-Fermi approximation. The results are shown in the left panel of Fig. \ref{fig:rho_mu_thomas_fermi}. 
At high densities, the self consistent $\rho$ vs $\mu$ curve agrees well with the
Thomas-Fermi approximation. However, at low densities (less than $5\times10^{-6}$/cell)  
the self-consistent scheme predicts a small residual charge in
the depletion zone that is absent in the Thomas-Fermi approximation. To shed light on the origin
of the discrepancy, we show the self-consistent $\rho_{conduction}$ vs $\mu$ relation 
for the conduction band electrons, without  contributions from the valence band.  The result, shown in the right panel of Fig. \ref{fig:rho_mu_thomas_fermi}, indicates a  better agreement with  Thomas-Fermi at low
densities. We conclude that, in addition to the surface contribution due to the characteristic helical metal,  the Thomas-Fermi approximation cannot capture the physics of the depletion zone, which is dominated by valence band contributions. In essence, this discrepancy is a quantum effect and is due to  the change in the valence state profiles induced by the nonuniform effective potential near the surface. Taking this effect into account may be important for the understanding of surface phenomena and for interpreting surface-sensitive measurements.

\subsection{Band-bending}

\noindent Understanding band bending in doped TIs is critical for
explaining the differences\cite{Analytis2010} between surface-sensitive
probes, such as the angle-resolved photoemission spectroscopy (ARPES),
and bulk-sensitive probes, for example Shubnikov--de Haas measurements.
In general, band bending in the vicinity of the surface is present
in all doped TI systems. However, surface dopants, such as Se vacancies\cite{Xia2009,Hsieh2009}
that migrate to the surface, can significantly modify the bending
of the bands at the boundary. Note that the concentration of carriers
generated by surface Se vacancies can be controlled by depositing
$O_{2}$ on the surface, which has been demonstrated\cite{Chen2010a}
to be an electron acceptor on Bi$_{2}$Se$_{3}$ surfaces.

\paragraph{Electron vs hole-doped topological insulators.}

\begin{figure}
%
%
%
\begin{tikzpicture}

\definecolor{mycolor1}{rgb}{0,0.5,0.9}
\definecolor{mycolor2}{rgb}{0.9,0.1,0.1}

\begin{axis}[%
xmin=1, xmax=30,
ymin=-0.0001, ymax=0.0001,
ylabel={$\rho_l$},
legend style={at={(0.97,0.03)},anchor=south east,nodes=right}]

\addplot [color=mycolor1,dashed,mark size=1pt,mark=*,mark options={solid}]
coordinates{
 (1,0.000977289307409368)(2,-5.60347460747401e-05)(3,-7.05455272300703e-05)(4,-7.28493743835656e-05)(5,-7.29196174118485e-05)(6,-7.2715100409404e-05)(7,-7.20813158809896e-05)(8,-7.05196313339762e-05)(9,-6.73124672050385e-05)(10,-6.18766283593253e-05)(11,-5.43474749918893e-05)(12,-4.58632049609164e-05)(13,-3.7979711453584e-05)(14,-3.15502254255584e-05)(15,-2.62845191727479e-05)(16,-2.15301164052661e-05)(17,-1.71961339088256e-05)(18,-1.36196890232476e-05)(19,-1.08455945349029e-05)(20,-8.54425070428988e-06)(21,-6.52456362404408e-06)(22,-4.87595581910583e-06)(23,-3.62919986862664e-06)(24,-2.63586107651648e-06)(25,-1.79396745325633e-06)(26,-1.1343550014864e-06)(27,-6.72523354783294e-07)(28,-3.49991964743168e-07)(29,-1.22921009371169e-07)(30,0)(31,-7.54951656745106e-15) 
};

\addlegendentry{electron doping};

\addplot [color=mycolor2,solid,mark size=1pt,mark=*,mark options={solid}]
coordinates{
 (1,-0.00084600595655715)(2,3.20759880199084e-05)(3,6.74826343558532e-05)(4,7.29176408544863e-05)(5,7.31573126453444e-05)(6,7.2308027951884e-05)(7,7.04001282869626e-05)(8,6.69580674104164e-05)(9,6.15893626445541e-05)(10,5.43434480173577e-05)(11,4.5907020682634e-05)(12,3.74459503424962e-05)(13,3.00930025571855e-05)(14,2.43942967979027e-05)(15,2.01292699131272e-05)(16,1.66474880418654e-05)(17,1.34345565050431e-05)(18,1.04362949189252e-05)(19,7.9193268485156e-06)(20,6.08050568029483e-06)(21,4.80974893513064e-06)(22,3.79928733207358e-06)(23,2.829862724818e-06)(24,1.9229291010614e-06)(25,1.23219807823816e-06)(26,8.27247361723238e-07)(27,6.0590051731424e-07)(28,4.06162885102646e-07)(29,1.73734142627779e-07)(30,0)(31,-1.55431223447522e-14) 
};

\addlegendentry{hole doping};

\end{axis}
\end{tikzpicture}

\caption{(color online) The local charge density for electron doped and hole doped TIs in the vicinity of the surface. The system has no surface dopants. The corresponding band bending near the surface has opposite directions for electron doped and hole doped systems and there is an approximate symmetry between the two cases.The
surface charge (corresponding to the first quintuple layer) is out of scale. \label{fig:charge_dopants_electron_hole}}
\end{figure}
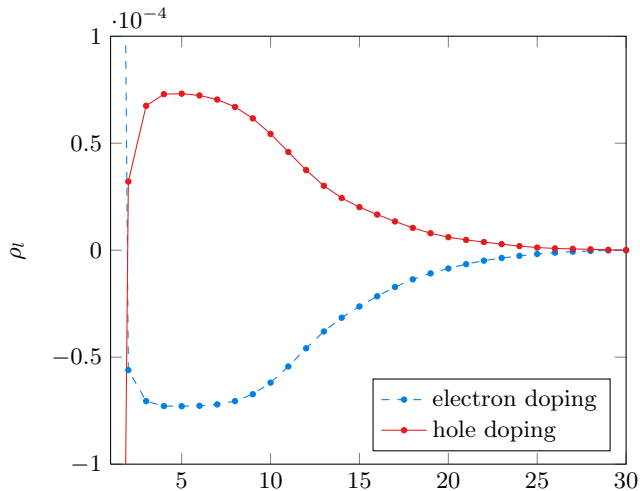

\begin{figure}
\centering

\subfigure[] {
\begin{tikzpicture} 
\begin{axis}[axis on top,width=0.5\columnwidth,height=2.7in,enlargelimits=false,xmin=0,xmax=40,ymin=-0.6,ymax=0.6,ylabel=Energy (eV)]
\addplot[thick,blue] graphics[xmin=0,ymin=-0.6,xmax=40,ymax=0.6] {edos_VN_E0\lyxdot 50002I0\lyxdot 50002R0L0_DV0_U04_Nz80};
\end{axis}
\end{tikzpicture}
\label{fig:ldos-nodopants-ndoped}
}\subfigure[] {
\begin{tikzpicture} 
\begin{axis}[axis on top,width=0.5\columnwidth,height=2.7in,enlargelimits=false,xmin=0,xmax=40,ymin=-0.6,ymax=0.6,label={(b)}]
\addplot[thick,blue] graphics[xmin=0,ymin=-0.6,xmax=40,ymax=0.6] {edos_VN_E0\lyxdot 49998I0\lyxdot 49998R0L0_DV0_U04_Nz80};
\end{axis}
\end{tikzpicture}
\label{fig:ldos-nodopants-pdoped}
}

\subfigure[] {
\begin{tikzpicture} 
\begin{axis}[axis on top,width=0.5\columnwidth,height=2.7in,enlargelimits=false,xmin=0,xmax=40,ymin=-0.6,ymax=0.6,ylabel=Energy (eV)]
\addplot[thick,blue] graphics[xmin=0,ymin=-0.6,xmax=40,ymax=0.6] {edos_VN_E0\lyxdot 50004I0\lyxdot 50002R0\lyxdot 000625L0\lyxdot 000625_DV0_U04_Nz80};
\end{axis}
\end{tikzpicture}
\label{fig:ldos-dopants-ndoped}
}\subfigure[]{
\begin{tikzpicture} 
\begin{axis}[axis on top,width=0.5\columnwidth,height=2.7in,enlargelimits=false,xmin=0,xmax=40,ymin=-0.6,ymax=0.6]
\addplot[thick,blue] graphics[xmin=0,ymin=-0.6,xmax=40,ymax=0.6] {edos_VN_E0\lyxdot 5I0\lyxdot 49998R0\lyxdot 000625L0\lyxdot 000625_DV0_U04_Nz80};
\end{axis}
\end{tikzpicture}
\label{fig:ldos-dopants-pdoped}
}

\caption{(color online) Local density of states 
as a function of the distance from the surface in  electron-doped (panels a and c) and hole-doped (panels b and d) TIs for $8\times 10^{-5}$ dopants/cell. The top panels (a and b) show the band bending in systems with no surface dopants, while the bottom panels (c and d) correspond to a system with $2.5\times 10^{-3}$ electron donor impurities on the surface. Note that the surface donors almost annihilate band bending in the n-doped system (panel c) and strongly enhances the effect in the p-doped TI (panel d). Adding electron acceptors on the surface will have an opposite effect.
\label{fig:ldos}
}
\end{figure}
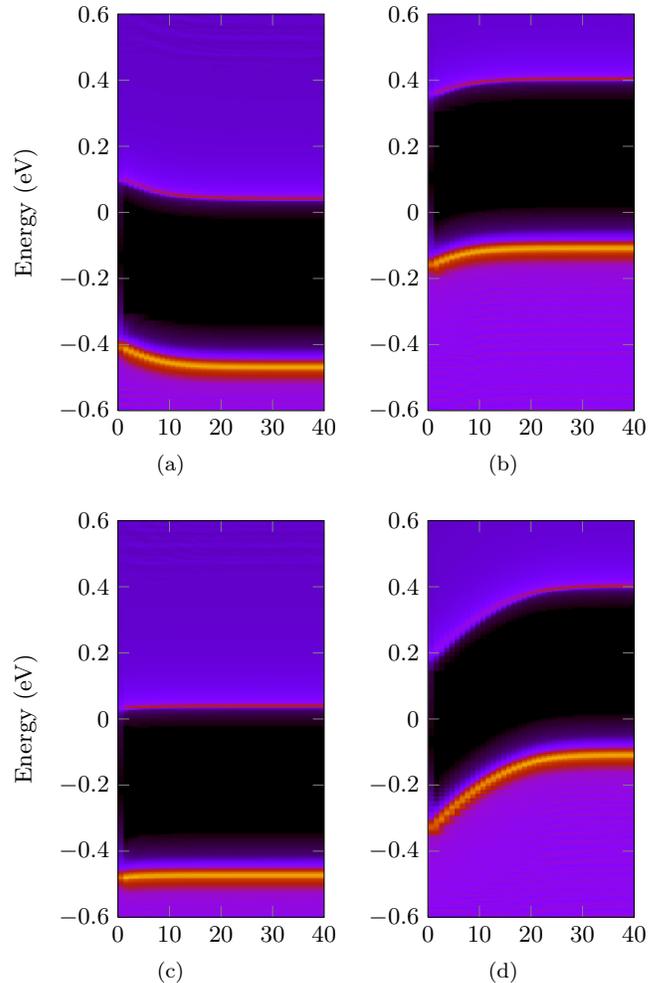

In the absence of surface dopants, the surface charge density profile exhibits an approximate 
electron-hole symmetry, as shown in Fig. \ref{fig:charge_dopants_electron_hole}. The depletion zone described above represents a region with lower carrier concentration, which corresponds to a lower electron density in n-type doped TIs and a higher electron density in p-type systems (see Fig. \ref{fig:charge_dopants_electron_hole}). 
These highly nonuniform density profiles  can be understood in terms of band bending in the vicinity of the TI surface. Rigorously speaking, band bending can be associated with a position-dependent local density of states (LDOS). The LDOS in the vicinity of the TI surface for both electron doped and hole doped systems is shown in Fig.~\ref{fig:ldos}. The approximate electron-hole symmetry that characterizes the system in the absence of surface dopants is manifest in 
in Fig. \ref{fig:ldos-nodopants-ndoped} and \ref{fig:ldos-nodopants-pdoped}. Quantitatively, in electron-doped systems with carrier densities of the order of $10^{18}cm^{-3}$ the band bending near the surface is positive   
and is characterized by an energy shift of about $60meV$. The p-doped case is characterized by a negative band bending of similar magnitude.

In the presence of surface dopants, the approximate electron-hole symmetry is broken. In particular,
as shown in Fig. \ref{fig:ldos-dopants-ndoped}  and \ref{fig:ldos-dopants-pdoped}, positively charged surface dopants (i.e., donors) strongly 
push the direction of the band bending downwards. As seen in the left panel of the figure, an appropriate
density of surface donors can basically eliminate band bending in an electron doped. By contrast, adding donors on the surface of a p-doped TI will further enhance the downward band bending (see Fig. \ref{fig:ldos-dopants-pdoped}). Of course, adding electron acceptors on the surface will have opposite effects, i.e., will enhance band banding in electron-doped systems and reduce or even reverse the effect in hole-doped TIs.

\paragraph*{Density profiles in the vicinity of the surface of a topological
insulator.}

\begin{figure}
\input{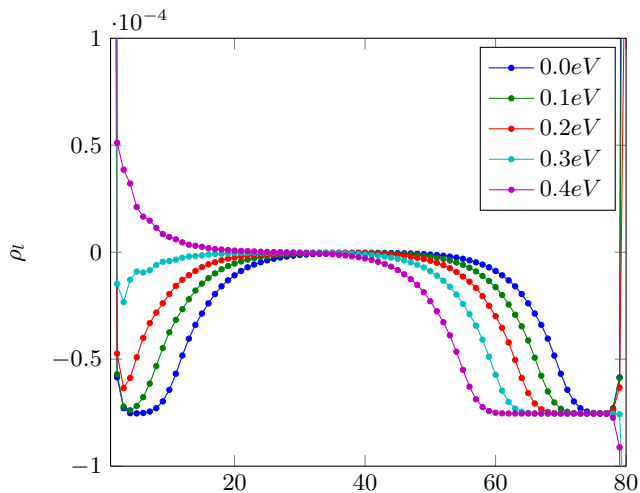}

\caption{(color online) Charge density profiles for an 80 layer n-doped TI slab with $8\times10^{-5}$ carriers per unit cell and no surface dopants. The different lines correspond to different gate voltage $\Delta V$. Note that, increasing the system size will result in expanding  the bulk region characterized by a constant density $\rho=0$, but the density profiles in the surface and the depletion (excess) regions will not be affected. Applying a negative gate voltage results in the expansion of the depletion zone, while a strong positive potential can creates a region with excess carriers near the surface. The profile of the transition zone (i.e., the region between the depletion region with no carriers and the bulk region with constant density) is independent on the applied potential, as long as depletion/bulk regions exist,  which is consistent with the Thomas-Fermi approximation. \label{fig:charge_gate_voltage_nodopants}}
\end{figure}

\noindent The intrinsically non-homogeneous density profiles induced by the surface charges generate
an effective electrostatic potential that varies significantly in
the vicinity of the TI surface. A powerful method to probe this potential
is the second harmonic generation (SHG) using ultrafast laser pulses.\cite{Hsieh2011,Hsieh2011a}.
Furthermore, applying gate voltages, which can allow the control of the charge carrier density in the vicinity of the surface, generates external fields that are nonzero in the depletion and surface regions. To determine the effect of a gate potential on the TI carrier density, we determine self-consistently the density profiles for different values of the gate potential. The results are shown in  Fig. \ref{fig:charge_gate_voltage_nodopants}. 
The calculations were done on an 80 layer TI slab. However, if a bulk region characterized by constant charge density (i.e., a neutral region) exists, the density profiles outside the bulk region are independent on the size of the system. Hence, our result are relevant for bulk TI crystals with $(1 1 1)$ surfaces.   Applying a negative gate voltage results in the expansion of the depletion zone. However, we find that the profile of the transition zone that separates the depletion region with no carriers from the bulk does not depend on the 
 applied potential.  This is consistent with the predictions of Thomas-Fermi approximation, which is not surprising if we consider our findings concerning the validity of this approximation in the transition regime  (see Fig. \ref{fig:rho_mu_thomas_fermi}). Finally, we note that a positive gate potential reduces the depletion zone and, if strong enough, can even create a region with excess carriers  near the surface.

\begin{figure}

\input{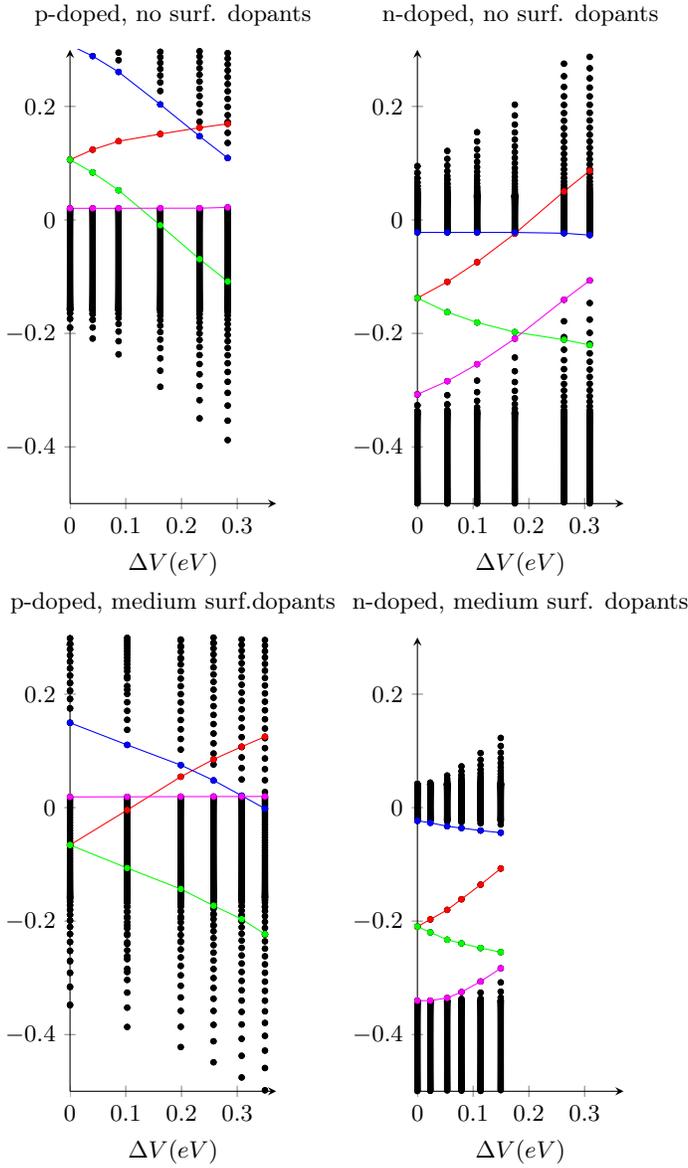}

\caption{(color online) Energy of the Dirac points and gap edge energies as functions of the applied gate potential.
To help the eye,  we draw lines corresponding to: the bottom of the conduction band (blue), the top of the valence band (magenta) and the upper (red)
and lower (green) Dirac pints. We note that, although the exact position of the Dirac cones relative to the bulk bands is model dependent, the general trends shown here are expected to hold in general. 
\label{fig:spectrum_voltage}}
\end{figure}

Next, we address the question concerning the dependence of the energy of the Dirac points and the dependence of the gap edges on the applied gate potential. The  results for both n-doped and p-doped TIs, with and without surface dopants, are shown in  Fig. \ref{fig:spectrum_voltage}. As expected, the effective gap is reduced by the external potential. The effect is also dependent on the concentration of surface dopants. Note that, depending on the type of bulk and surface dopants, as well as the applied gate potential, the Dirac points  can be inside the bulk gap or buried inside either the valence or the conduction band. Using the right combination of dopants and applied gate potentials is key whenever exposing the Dirac point is critical for probing certain physical properties, e.g., the opening of a gap at the Dirac point.

\section{Electrostatic effects in topological insulator thin films}

In thin TI films, i.e., films containing $\sim10$ layers ($10nm$), most of the carriers may come from surface donors. Consequently, it is possible that the ``bulk'' is no longer charge neutral. Strictly speaking, in most thin TI films there is no bulk region as defined above, i.e., a neutral region characterized by a uniform charge density. Also, external potentials can create strong perturbations, as screening is significantly weaker than in bulk TIs. All these characteristics lead to certain specific electrostatic effects that are not present in bulk systems.

\subsection{Substrate induced band bending in thin films}

\noindent Thin films are often grown on substrates using, for example, molecular beam epitaxy (MBE).
In epitaxially grown films, a potential difference between the two surfaces
of the TI may occur as a result of the charge transfer from the substrate.\cite{Zhang2010} Furthermore, 
in the ultra-thin film limit (less than 6 quintuple layers) the overlap
between the surface states from the two surfaces of the film cannot
be neglected and, as a result, a gap opens at the Dirac point.
In the presence of a substrate-induced potential difference, an effective
Rashba type spin-orbit splitting is generated at ${\bf k}\neq0$\cite{Zhang2010}.

\begin{figure}
\includegraphics[width=1\columnwidth]{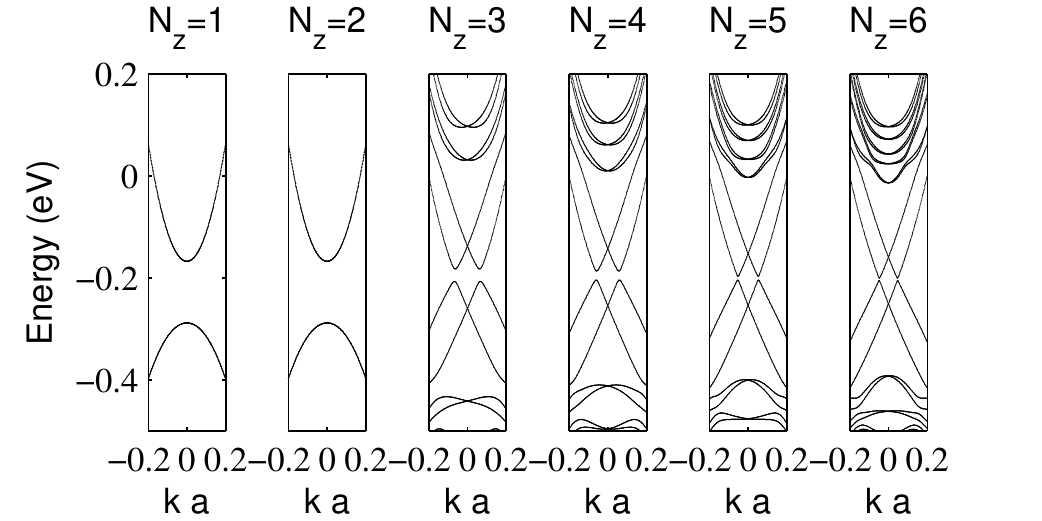}

\caption{Ultra thin film spectra in the presence of a substrate-induced potential. Note the 
Rashba type band splitting splitting for $N_z\geq 3$.  Because the surface states extent about three layers away from the film surface, they hybridize with the surface states on the opposite side, which gives rise to a gap.
\label{fig:Rashba-splitting}}
\end{figure}

First, we study a doped ultra thin TI film in the presence of a substrate-induced potential difference. The spectra  for films with six quintuple layers or less are shown in Fig. \ref{fig:Rashba-splitting}. Note that the Rashba-type band spitting and the gap opening  in the surface spectrum due to the overlap of the surface states localized near the two surfaces  are in good agreement with the experimental findings~\cite{Zhang2010}. 

\begin{figure}

\subfigure[]{
\begin{tikzpicture} 
\begin{axis}[axis on top,width=0.41\columnwidth,height=3in,enlargelimits=false,xmin=0,xmax=9,ymin=-0.6,ymax=0.6,ylabel=Energy (eV)]
\addplot[thick,blue] graphics[xmin=0,ymin=-0.6,xmax=9,ymax=0.6] {edos_VN_E0\lyxdot 5I0\lyxdot 5R0L0_DV0\lyxdot 1_U04_Nz10};
\end{axis}
\end{tikzpicture}
}\subfigure[]{
\begin{tikzpicture} 
\begin{axis}[axis on top,width=0.41\columnwidth,height=3in,enlargelimits=false,xmin=0,xmax=9,ymin=-0.6,ymax=0.6,ytick={\empty}]
\addplot[thick,blue] graphics[xmin=0,ymin=-0.6,xmax=9,ymax=0.6] {edos_VN_E0\lyxdot 50025I0\lyxdot 50025R0L0_DV0\lyxdot 1_U04_Nz10};
\end{axis}
\end{tikzpicture}
}\subfigure[]{
\begin{tikzpicture}
\begin{axis}[axis on top,width=0.41\columnwidth,height=3in,enlargelimits=false,xmin=0,xmax=9,ymin=-0.6,ymax=0.6,ytick={\empty}]
\addplot[thick,blue] graphics[xmin=0,ymin=-0.6,xmax=9,ymax=0.6] {edos_VN_E0\lyxdot 50125I0\lyxdot 50125R0L0_DV0\lyxdot 1_U04_Nz10};
\end{axis}
\end{tikzpicture}
}

\caption{(color online) Local density of states for a 10 layer TI film in the presence of an external potential difference, $\Delta V=0.1eV$. (a) $\delta n_e=0$, (b) $\delta n_e=10^{-3}/({\rm unit cell})$
(c) $\delta n_e=5\times10^{-3}/({\rm unit cell})$. Note that the band bending effect depends strongly on the carrier concentration. 
\label{fig:ldos_nz10}}
\end{figure}
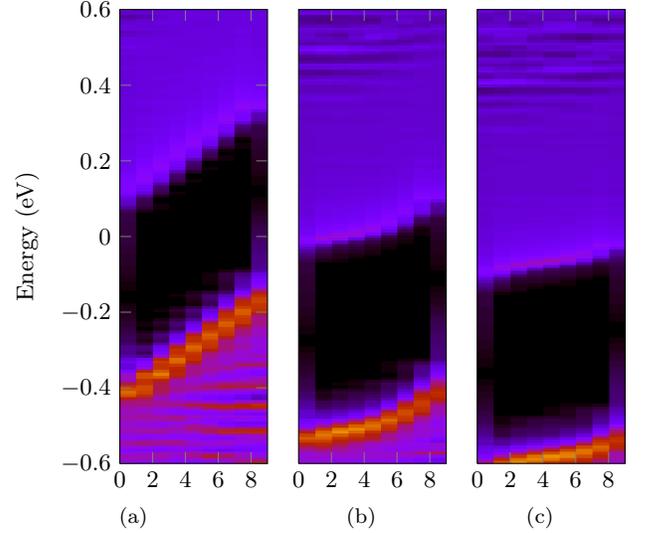

\begin{figure}

\begin{tikzpicture}

\definecolor{mycolor1}{rgb}{0,0.75,0.75}

\begin{axis}[ylabel={$\left|\psi_l\right|^2$}]

\addplot [color=blue,solid,mark=*,mark options={solid}]
coordinates{
 (1,0.921042820747996)(2,0.066997457932844)(3,0.0102253114135455)(4,0.00144161336832109)(5,0.000232818811722223)(6,4.22409502567242e-05)(7,8.93498404570302e-06)(8,2.30735575167537e-06)(9,7.89843182049513e-07)(10,3.95116846068448e-07)(11,2.86912617863914e-07) 
};
\addlegendentry{Surface};

\addplot [color=green!50!black,solid,mark=*,mark options={solid}]
coordinates{
 (1,0.283298449874217)(2,0.362353456857594)(3,0.235790941417686)(4,0.0895228135499084)(5,0.0232874358813779)(6,0.00477362962704565)(7,0.000826187516690811)(8,0.000126657362088051)(9,1.77623927277003e-05)(10,2.33322494388872e-06)(11,2.92198183383741e-07) 
};

\addlegendentry{Conduction n=1};

\addplot [color=red,solid,mark=*,mark options={solid}]
coordinates{
 (1,0.280952937055801)(2,0.362787330648732)(3,0.236943908126942)(4,0.0900872749242358)(5,0.0234432788902272)(6,0.00480567965687899)(7,0.000831592418542236)(8,0.000127450675893763)(9,1.7867507789734e-05)(10,2.34612255229546e-06)(11,2.93690350405419e-07) 
};
\addlegendentry{Conduction n=2};

\addplot [color=mycolor1,solid,mark=*,mark options={solid}]
coordinates{
 (1,0.183828394162165)(2,0.283802605099121)(3,0.160293834052975)(4,0.199675979605588)(5,0.117704128677183)(6,0.0414444603560234)(7,0.0106116130016736)(8,0.00218380908736639)(9,0.000384676961622623)(10,6.04314765319396e-05)(11,8.71553062109249e-06) 
};
\addlegendentry{Conduction n=3};
    
\end{axis}
\end{tikzpicture}
\caption{Profiles of surface and bulk states for the same parameters as
in Fig.\ref{fig:ldos-dopants-pdoped}. In this case the Dirac point is buried inside the valence but the surface Fermi line is exposed due to band bending. The substantial spatial 
overlap  may allow optical transitions between occupied surface states and empty conduction states. 
\label{fig:psi_profiles}}
\end{figure}
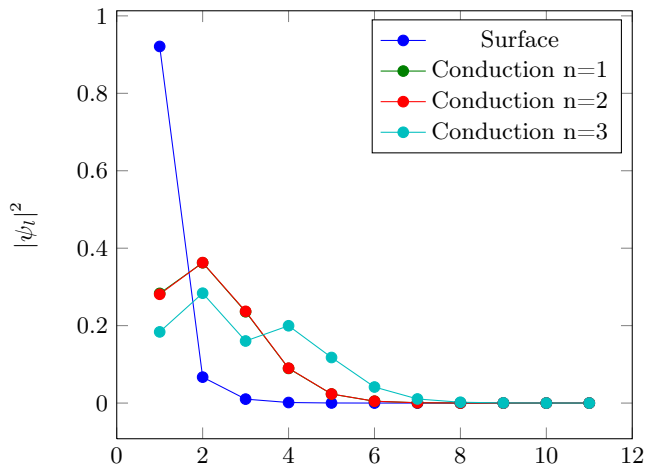

Next, we address the question of band bending in thin films. As mentioned above, in thin films the external fields are weakly screened. This provides an effective way to control the chemical potential in doped
TI thin films and to realize bulk-insulating systems by applying
gate voltages to remove the excess bulk charge carriers.\cite{Chen2010,Kim2011}
We note that, in systems characterized by weak substrate-induced band
bending, it is possible to have the Dirac points associated with the
two surfaces approximately at the same energy\cite{Kim2011}. After removing the excess carriers, the film is no longer charge neutral and an effective potential with a parabolic layer dependence  is created inside the film. Nonetheless, for typical film sizes and initial carrier concentrations [e.g. 10 layers, $8 \times 10^{-5}$ electrons/cell] the corresponding band bending effects are negligible. For comparison, we note that, for a TI film with an initial carrier concentration of $8 \times 10^{-5}$, after completely depleting the excess charge the gap will close due to band bending effects if the thickness of the film exceeds 60 layers. 

In addition to the effective potential generated by the removal of charge carriers, external gate potentials and substrate-induced potential differences can significantly modify the spectrum. 
In Fig. \ref{fig:ldos_nz10} we show the self-consistently calculated band bending of a thin film with
no surface dopants for three different values of the carrier density.  Note that, for a given potential difference, the intensity of the band bending effect depends strongly on the carrier density. As expected, higher carrier concentrations 
provide better screening, which results in a weaker band bending. Also note that, depending on the carrier density, the chemical potential may be inside the gap (for films with a density close to half filling), it may cross the bottom of the  conduction band  at intermediate doping,  or it may be completely inside the conduction band
at large filling.\cite{Chen2010}

\subsection{Surface states and bulk states in topological insulator films}

In TIs the surface states associated with $(111)$ surfaces are localized
within about one quintuple layer from the surface and, consequently,
the matrix elements of any operator between surface and bulk states
vanish in the thermodynamic limit. However, in thin films the overlap
between surface and bulk states becomes finite. Moreover, we show
that, in the presence of a local electric field that pushes the electrons
toward the surface, the lowest energy bulk states become localized
in the vicinity of the TI surface. This opens the possibility of optically
coupling surface and bulk states in topological insulators with finely
tuned surface and bulk density. One possibility is to use a p-doped TI with strong downward band bending near the surface (See Fig.\ref{fig:ldos-dopants-pdoped}). The tip of the Dirac cone will be buried inside the valence band, but the Fermi line corresponding to the surface states will be exposed due to  the strong band bending. For any given wave vector in the vicinity of the surface Fermi k--vector, there are empty conduction states above the occupied surface states. These surface and bulk-type states could be coupled optically, if the corresponding matrix elements are nonzero. Typically, the main problem in TIs is that the surface states are localized within a few quintuple layers from the surface while the bulk states are extended, hence the overlap is vanishingly small in the thermodynamic limit. However, in the presence of strong negative band bending, the low-energy conduction states become localized near the surface. This situation, which ensures the necessary condition for a strong surface-bulk overlap, is illustrated in Fig. ~\ref{fig:psi_profiles}.  Future studies are necessary to determine explicitly the optical matrix elements  associated with transitions from the surface states to the conduction band in doped TIs with strong band bending. 

\section{Summary and conclusions}

We use a self consistent scheme to study electrostatic effects in
topological insulators. We find that in the vicinity of the TI surface
a depletion zone characterized by the absence of bulk carriers may occur.
We find that the Thomas-Fermi approximation fails in this depletion zone and
in the surface region where the contribution from the topological states is dominant.
As a result of the intrinsic charge inhomogeneity in the vicinity of the surface
we find that band bending effects are typically present. In the absence of surface
dopants the band bending for electron and hole systems have similar magnitudes
and opposite sign. The presence of surface dopants breaks this
approximate electron-hole symmetry and dramatically affects the magnitude
of the effect. We find that applying an external gate voltage can expand
the depletion zone away from the surface. The transition region between the
depletion zone and the bulk is characterized by a charge profile that is
independent of the applied potential, in accordance with the Thomas-Fermi approximation.
In thin films there is less bulk charge and, consequently weaker screening effects are observed.
We show that band bending due to external fields depends strongly on the
total charge density of the sample. We find that substrate induced potentials 
can generate a Rashba type spin-orbit coupling in ultra thin TI films.
We calculate the surface and bulk state profiles and find that, in certain conditions, they 
are localized within the same region in space. Future work is needed determine
if selection rules allow optical transitions between these states.

\begin{acknowledgments}
We would like to thank Dennis Drew for stimulating and useful discussions.
\end{acknowledgments}
\bibliography{BBendingBib}

\end{document}